\documentclass[pra,twocolumn,twoside,aps]{revtex4-1}

\usepackage{amsmath, amssymb}
\usepackage{bm}
\usepackage{natbib}
\usepackage{listings}
\usepackage{url}
\usepackage[dvipsnames]{xcolor}
\usepackage{soul}
\usepackage{enumitem}
\usepackage{amsmath, amssymb}
\usepackage{bm}
\usepackage{natbib}

\input{listings-python.prf}
\usepackage{url}
\usepackage[dvipsnames]{xcolor}
\usepackage{setspace}
\usepackage{graphicx}
\usepackage{float}

\begin{document}

\title{Transmission of natural scene images through a multimode fibre}


\author{Piergiorgio Caramazza\textsuperscript{1}, Ois{\'i}n Moran\textsuperscript{2},  Roderick Murray-Smith\textsuperscript{2}, Daniele Faccio\textsuperscript{1} }

\affiliation{$^1$School of Physics and Astronomy, University of Glasgow, Glasgow, G12 8QQ, UK\\
$^2$School of Computing Science, University of Glasgow, Glasgow, G12 8QQ, UK}
\email{Daniele.Faccio@glasgow.ac.uk,\\ 
Roderick.Murray-Smith@glasgow.ac.uk}

\begin{abstract}
\noindent {\large{\bf{Abstract }}} \\
The optical transport of images through a multimode fibre remains an outstanding challenge with applications ranging from optical communications to neuro-imaging. State of the art approaches  either involve measurement and control of the full complex field transmitted through the fibre or, more recently, training of artificial neural networks that however, are typically limited to image classes belong to the same class as the training data set.
Here we implement a method that statistically reconstructs the inverse transformation matrix for the fibre. We demonstrate imaging at high frame rates, high resolutions and in full colour of natural scenes, thus demonstrating general-purpose imaging capability. Real-time imaging over long fibre lengths opens alternative routes to exploitation for example for secure communication systems, novel remote imaging devices, quantum state control processing and endoscopy. 

\end{abstract}

\maketitle

\noindent {\large{\bf{Introduction }}} \\
\indent Optical fibres form the backbone of the internet and are a key technology in modern society \cite{Hecht2}. The vast majority of these fibres are `single mode' i.e. they can transmit only one single, roughly Gaussian-shaped beam profile, corresponding to the so-called fundamental mode of the fibre \cite{Hecht}. It is therefore impossible to directly transmit images through an optical fibre: any attempt to do so simply results in transmission of the one single allowed mode and therefore the detection at the output of this (Gaussian-shaped) mode with all other information of the image completely lost at the fibre input. 
One possibility to circumvent this limitation is to resort to an array or bundle of single mode optical fibres, each one transmitting the information of a single pixel in the output image. However, this quickly leads to fibre-bundled cables that are relatively thick and not optimal for applications such as endoscopic or neurological imaging, where the fibre bundle is inserted inside a body \cite{fibre_review}. \\
Another option is to resort to multimode fibres, i.e. fibres that due to a larger core diameter can carry many optical modes that will have more complex shapes than the fundamental mode and may encode image information \cite{cizmar3}. For example, a typical 100 $\mu$m core diameter fibre might carry around 10,000 modes  and could in principle transmit an image with roughly the same number of pixels. However, in these fibres each of these individual modes propagates at a slightly different velocity, thus leading to an amplitude and phase mixing of the image as this propagates along the fibre \cite{fibre_book}. The image at the fibre output therefore appears as a random array of bright and dark spots, referred to as a speckle pattern. This effect cannot be avoided and completely scrambles and destroys the input image. Full a priori knowledge of the input image and fibre details could allow to numerically model the optical propagation \cite{theoryMMF}, reconstruct the transmission matrix and then unscramble the output data, but in practice this can be extremely hard. Methods have been developed that allow to shape the input beam profile so as to focus the output field into a single spot that can then be scanned \cite{MMF_holo,MMF_holo2,MMF_holo3,MMF_holo4,ploschner2014gpu}  with an emphasis on endoscopy \cite{choi2012scanner,cizmar1,cizmar,cizmar_brain}. Notwithstanding this notable progress, the development of a viable method that allows to unscramble the speckle patterns and thus retrieve high resolution, general image information in real time is an open challenge.\\
 \begin{figure}[t!]
	\centerline{\includegraphics[width=8cm]{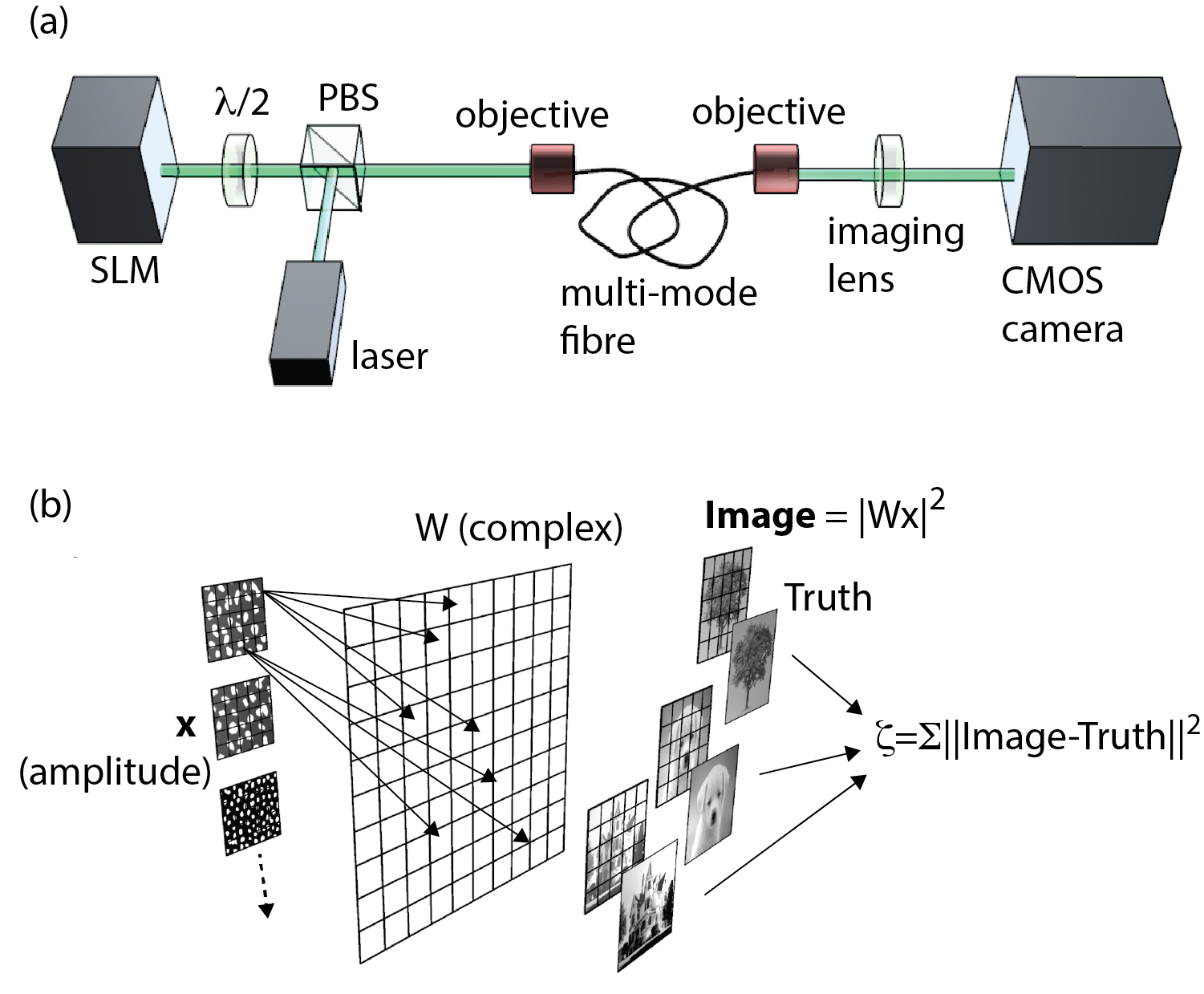}}
	\caption{ Experimental layout. (a) An SLM combined with a polarising beam splitter (PBS) and a half-wave plate ($\lambda/2$) is used to imprint intensity images onto a laser beam that is then coupled into a multimode fibre. The fibre output is collected with a lens and recorded on a CMOS camera. (b) A schematic overview of the computational processing steps {of the inversion process}: output speckle data, $x$ from a series of images (in our experiments, 50,000 images from the ImageNet database {\cite{imagenet_cvpr09}}) are fully connected to a complex matrix, $W$ which provides an output image $I=|Wx|^2$. This image is compared to the actual original image (ground truth) through a cost function: the total cost $\zeta$ is then back-propagated to W and the process is repeated for a fixed number of loops (epochs), ensuring minimisation of $\zeta$.}
	\label{fig:layout}
\end{figure}
One promising route in this direction is based on the complete characterisation of the optical fibre in the form of a measurement of its transmission matrix \cite{choi2012scanner,cizmar3,cizmar}. This matrix connects certain orthogonal modes at the fibre input to the fibre output and can therefore, once known, be used to invert the speckle pattern back into the original image. This approach requires measurements of the full complex (amplitude and phase) profile of a large subset of modes and has been shown to work over fibre lengths of 0.3-1 m. \\
 Other approaches pioneered by Takagi et al. \cite{Hori}, have recently been proposed that used artificial neural networks (ANNs) {using deep learning encoders} to infer images from the speckles patterns without any need for an a priori mathematical model of the fibre \cite{psaltis1,psaltis2,Su_fibre}. These have used multi-layer convolutional ANN's and have shown that it is possible to reconstruct hand-written digits from the MNIST database {\cite{lecun98}} that consists of patterns with $28 \times 28$ pixel resolution: training of the network and testing of its reconstruction abilities are both performed on digits from the same database. ANNs have also been shown to allow for example to focus a beam through a thin scattering medium (as well as through a multimode fiber) in \cite{alex}, where both single layer and multi-layer real-valued neural networks have been implemented.  When used for imaging, these approaches are mostly expected to work for classes of objects that belong to the same class used for the training as explicitly pointed out by Psaltis et al. \cite{psaltis1}. First steps towards generic imaging have been made in Ref.~\cite{psaltis2}: these approaches present a very promising route forward if the suitability for general purpose imaging applications can be addressed.  \\
 \begin{figure}[tb!]
	\centering
	\includegraphics[width=8cm]{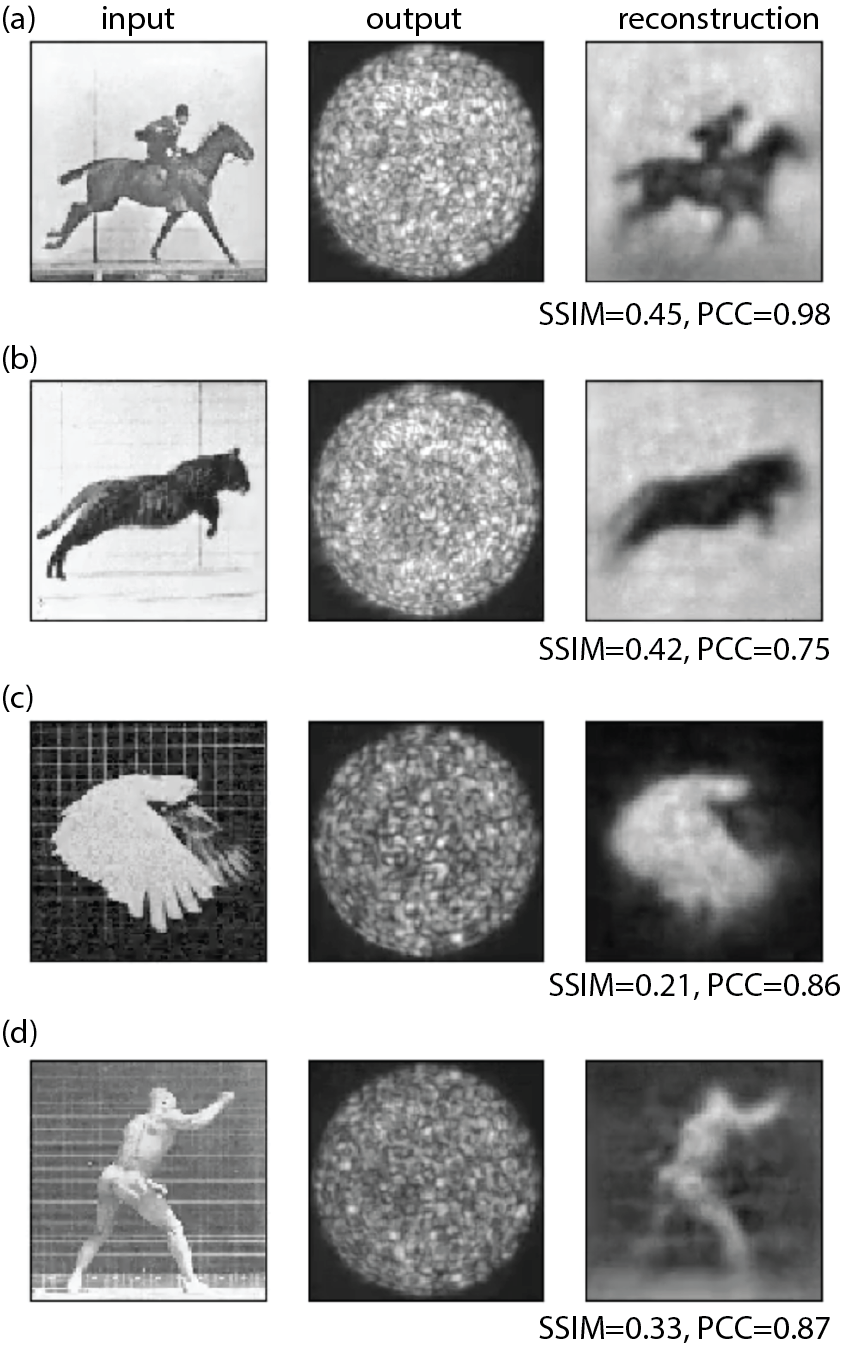}
	\caption{Reconstruction of Muybridge videos for 1 m long fibre. Individual frames of video data taken from the Muybridge collection: a) running horse, b) jumping cat, c) flying parrot, d) punching boxer. All videos are grayscale, scaled to $92 \times 92$ pixels and transmitted at 4 frames/second. The first column shows the original input video. The second column shows the speckle patterns ($x$) measured at the fibre output. SSIM and PCC indicate the structural similarity index and Pearson correlation coefficient that quantify the quality of the reconstruction (see Supplementary Note 2).}
	\label{fig:results1}
\end{figure}
We have developed an approach that allows us to transmit and reconstruct images of natural scenes at high resolution and frame rates. Our method resorts to building an approximate model of the inverse of a complex-valued, intensity transmission matrix of the optical fibre. This approach does not require the use of deep (multi-layer) ANNs and enables the full reconstruction of detailed images.    Full colour images and videos can be recorded at 20 fps and could be scaled up to kfps.  \\
\\
\noindent{\large{\bf{Results}}}\\
\indent {\bf{Experiments.}} The experimental layout is shown in Fig.~\ref{fig:layout}(a). We use an SLM (maximum frame rate of 20 Hz) to impart grayscale (100 grayscale levels) intensity images on to a continuous-wave laser beam (532 nm wavelength). This image is then coupled into a multimode fibre ({step index core, core diameter {105} $\mu$m}, fibre lengths of 1 m and 10 m, $\sim9000$ propagating optical modes, image spot size at fibre input is $\sim2$ $\mu$m) and then coupled out using identical objectives for the fibre input and output (focal length, $f=34$ mm, NA$ = 0.26$). The speckle pattern at the fibre output (near field)  is imaged on to a CMOS camera at $350\times350$ pixel resolution. \\
The goal is to transmit `natural scenes', i.e.  photographs of everyday-life scenes. The importance of this choice lies in the significant additional complexity of natural scenes when compared to e.g. MNIST-database digits or other simple geometric features.  As sample images we use a selection of 50,000 images from the Imagenet database \cite{imagenet_cvpr09}, sized at $92\times92$ pixels so as to have less pixels than optical fibre modes. These images have been selected randomly from the Imagenet database while looking for almost square images so as to facilitate projection into the fibre. 
The output speckle patterns with amplitude distribution, $x$ (i.e. $x$ is the square root of the measure speckle intensity patterns), together with the knowledge of the image (intensity distribution, $I$) that generated each speckle pattern, is used in the algorithm described below to approximate the inverse of a complex transmission matrix, $W$. This matrix is then used to retrieve images that were not part of the sample data set from intensity measurements of their output speckle patterns, $I=|Wx|^2$. Examples are images and videos from from the Muybridge collection such as a running horse, a jumping cat and a flying parrot. We also tested the imaging on videos of a rotating Earth and Jupiter. Both of these are in full colour, obtained by projecting and then recombining the R, G and B channels independently.\\
\indent{\bf{Image reconstruction.}}
  There are two possible approaches to reconstructing an image from a speckle pattern. The first is to attempt to build a forward model that describes how the images or optical modes propagate down the fibre and then invert this or, the approach followed here, one can try to directly construct an approximation of the inverse model. This is achieved statistically (i.e. by employing data from many images) through a single, fully-connected complex-valued transformation matrix.\\
    %
 \begin{figure}[t!]
	\centering
	\includegraphics[width=7cm]{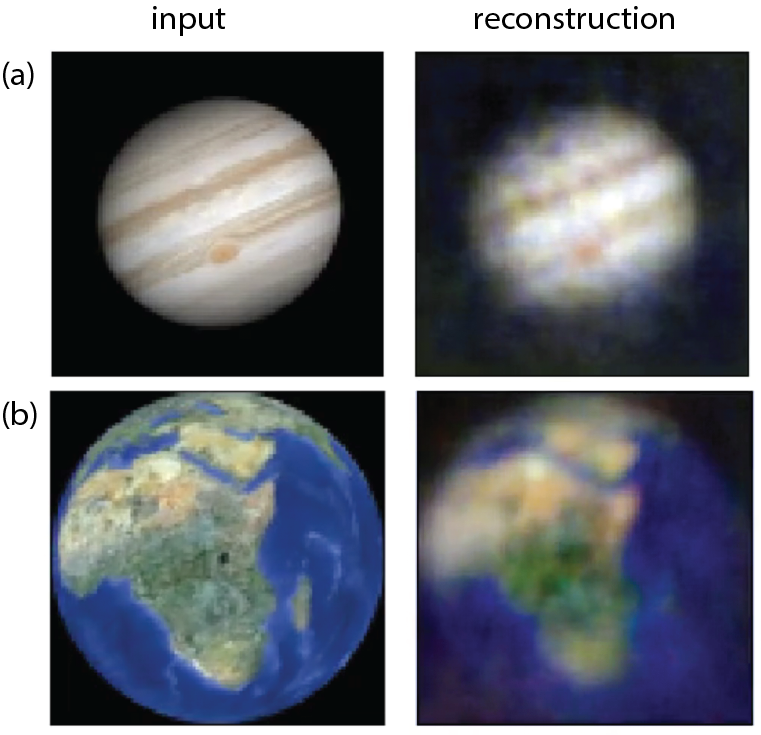}
	\caption{Full colour results for 1 m long fibre. Individual frames of video data in full colour of a rotating Jupiter (credit: Damian Peach) (a) and Earth (b).}
	\label{fig:results2}
\end{figure}
{\bf{Complex inversion.}} A schematic overview of the approach used to reconstruct $W$ is shown in Fig.~\ref{fig:layout}b). The full transmission matrix of an optical fibre is complex-valued. This motivates the assumption that $W$ is also complex valued, connecting the input and output  of the fibre $I=|Wx|^2$ \citep{Hir03a,TygBruChi16,Gub16,Tra18}. This also motivates idea that a deep-learning ANN approach is not required here.  We only measure the  intensity of the speckle pattern from which we take the amplitude $x$ (square root of the intensity)  and therefore {represent $x$ with amplitude only, and zero phase.} 
The values of $x$ {are passed to a fully connected (`dense') complex matrix -- equivalent to multiplication by the complex-valued matrix $W$} [arrows in Fig.~\ref{fig:layout}b) show some of these connections as an example]. An image is then obtained as {$I=|W{x}|^2$}. {We calculate the derivatives ${d\zeta}/{dw_{ij}}$ of the cost function $\zeta$, with respect to the $i,j$th element of $W$. We then apply a stochastic gradient descent approach to make small changes to $W$ that reduce the cost-function and the process is repeated for a fixed number of loops (epochs), ensuring convergence of $\zeta$ to a minimum value. The complex-weighted inversion was implemented as a novel layer with Keras \citep{chollet2015keras} and TensorFlow \citep{tensorflow2015-whitepaper}, {(the code is provided in the Supplementary Note 4)}. 
The final $W$, constructed from a database of 50,000 images can then be used to obtain an estimate of the ground truth image for all future transmitted data corresponding to images not used as part of the training and indeed, even transmitted at a completely different time (e.g. several days) after the training is completed.\\
\begin{figure}[tb!]
\includegraphics[width=8cm]{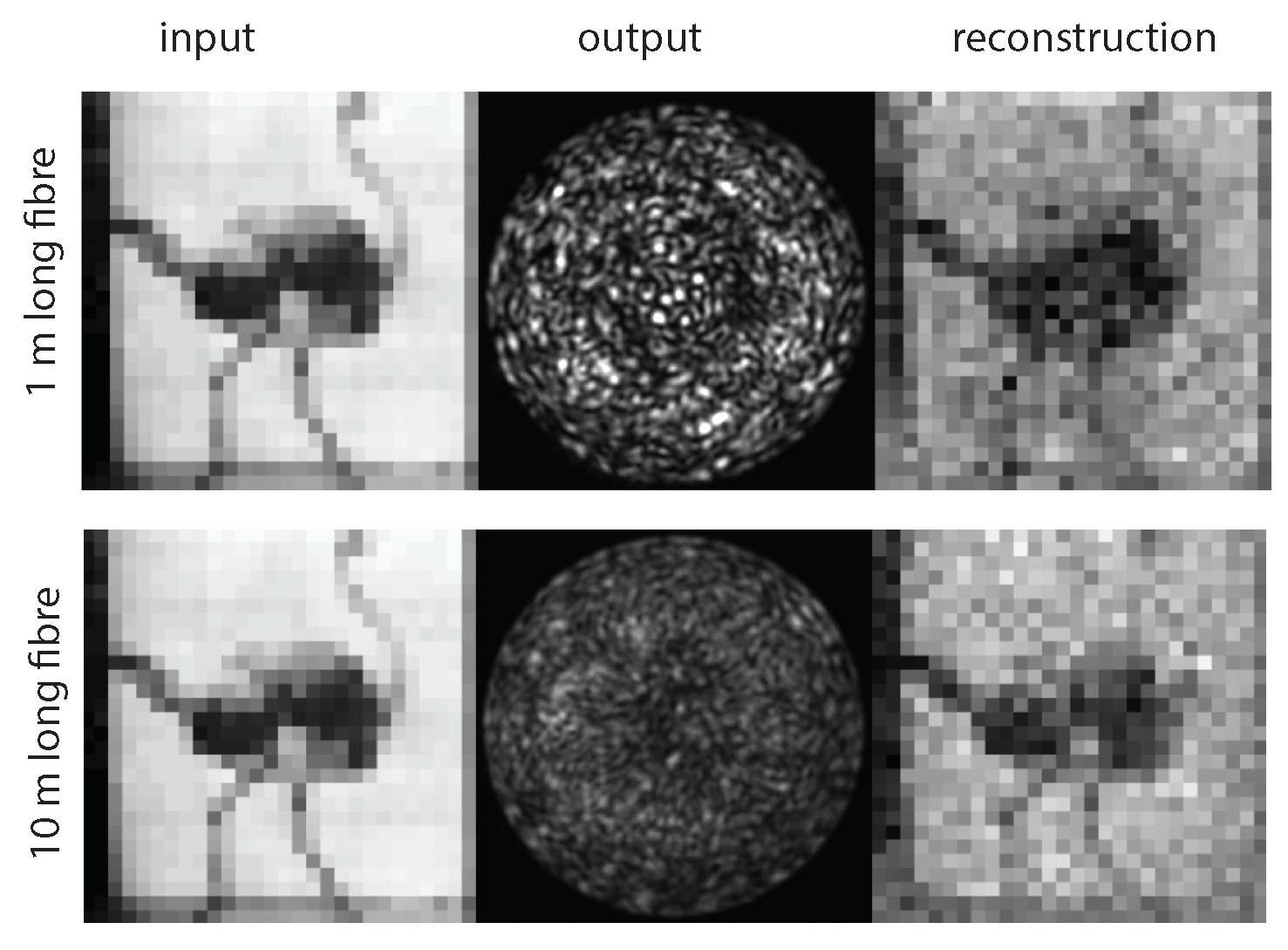}
\caption{Image reconstruction of an ostrich taken from the Muybridge collection after transmission through a 1 m and 10 m long fibre. The inversion matrix was constructed using only random grayscale patterns.}
\label{fig:ostrich}
\end{figure}
{\bf{Experimental Results.}} In Fig.~\ref{fig:results1} we show a first set of results obtained from data transmitted through a 1 m long fibre (wrapped in a loose coil on the table) by applying the {estimated} $W$ to a series of videos that do not form part of the ImageNet database and are significantly diverse, in order to demonstrate the robustness of our approach. These videos are taken from the Muybridge recordings from the 1870's that marked the historically important breakthrough of the first ever high-speed photography images. The running horse in Fig.~\ref{fig:results1}a) is probably the most iconic of the Muybridge videos, but the others, a jumping cat  in Fig.~\ref{fig:results1}b), a flying parrot  Fig.~\ref{fig:results1}c) and punching boxer  Fig.~\ref{fig:results1}d) provide a broad scene variability and all show good image reconstruction over the full gray-scale spectrum as opposed to the binary black/white MNIST images often used in previous work. {{We also note that the MNIST database, given the limited set of symbols, will tend to create what is essentially a simple classification system, with a decoder to generate the associated image. True imaging capability should therefore demonstrate functionality beyond this dataset, as shown in Fig.~\ref{fig:results1} (see Supplementary Figure 3 for futher examples).
   Under each reconstructed image in Fig.~\ref{fig:results1} we also give a quantitive measure of the reconstruction quality based on the `structural similarity index' (SSIM) and the `Pearson correlation coefficient'  (see Supplementary Note 2). These coefficients measure the similarity between the ground truth and retrieved images with a maximum value of 1 (indicating image identity).\\
   We also observed no degradation in the video quality even when the data was transmitted, recorded and reconstructed more than 48 hours after transmission of the original first set of `training' data had been completed (see Supplementary Note 3), thus indicating robustness to subsequent environmental changes such as temperature fluctuations (of the order of a few degrees) and vibrations (the setup is not placed on a vibration-isolated table). \\
%
%
In Fig.~\ref{fig:results2} we show examples of full colour video transmission of a rotating Jupiter and a rotating Earth. Each individual R, G and B channel was transmitted and reconstructed separately and then  recombined.  We note that the same matrix $W$ obtained for grayscale images is used for all three R, G, B channels when imaging in full colour mode. Subtle features such as the Red Spot on Jupiter or slightly lighter areas in the Northern region of Africa (roughly corresponding to the Nile delta region in Egypt) can be observed in the reconstructed images (other examples are shown in the Supplementary Fig. 2). \\
Tests were also performed to investigate the role of the class of images used for retrieving $W$.  The images in Fig.~\ref{fig:ostrich}  were down-sampled to $28 \times 28$ pixels in order to simplify the problem and demonstrate that, if desired, one may also reconstruct the inversion matrix $W$ with a completely ``agnostic'' approach, i.e. with no prior assumption on the images. This is obtained by using 50,000 completely random grayscale images. As can be seen, this completely agnostic approach is still able to correctly reconstruct the images although with a clear loss of quality. We noticed a good insensitivity to fibre length (as already pointed out by Psaltis et al. \cite{psaltis1}) and an improvement of image quality with increasing number of random images used for the $W$ matrix retrieval, although GPU RAM limitations did not allow us to investigate this further. \\
We also noted that changing the size of the focused image at the fibre input significantly impacts the final reconstruction. By placing a telescope after the SLM so as to rescale the image at the focusing objective input, we noticed a significant increase of the final image quality with increasing size at the focusing objective input pupil (corresponding to an increasing effective NA, i.e. to an increasing angular spread at the fiber input). This was also accompanied by a clear decrease in the average speckle spot size at the fibre output, indeed indicating the excitation of higher spatial frequency modes  (see Supplementary Note 1 and Supplementary Figure 1).\\
 A matter of concern in many studies is the robustness to changes in the fibre configuration. A change in the fibre geometry (e.g. by bending the fibre) will lead to a different  propagation of the individual modes, which ultimately leads to a different output speckle pattern. Without precise knowledge of how  the fibre has been changed, it is not possible therefore to reconstruct the image using the inversion matrix $W$ from a different configuration \cite{cizmar}. A recent solution has been proposed by using specially designed fibres that have a parabolic refractive index profile in the core \cite{cizmar2018}. 
For a long-range transmission system, such a solution might be appropriate under the assumption that a long-haul fibre would remain in a relatively fixed position over time.\\
In future realisations, we expect that a combination of fibre design, position classification and/or extensive training over fibre configurations, will allow to efficiently remove this last obstacle that for the time being, is beyond the scope of this work that is aimed at demonstrating that high pixel-density, colour images can be transmitted efficiently through a static fibre and at video frame rates.\\
%
\\
\noindent{\large{\bf{Discussion}}} \\
\indent Imaging through a single multimode fibre is an outstanding challenge and has so far been limited to relatively low frame rates, small image sizes or short fibre lengths. Imaging of natural scenes increases this challenge further as it ideally entails video frame rates, colour detail and sufficient image resolution to allow identification of the scene details.  The technique developed here is based on a physically informed model of the imaging system that retrieves an approximation to the full transmission matrix. In this sense, our approach sits somewhere between the techniques devised for reconstructing the actual transmission matrix and deep learning approaches that do not make any explicit assumptions on the system.  This allows for efficient video and data transmission through fibres and promises applications beyond endoscopic imaging, such as direct video or multimode data transmission over long fibres for communication systems and fibre sensing by, for example, exploiting the image sensitivity to changes along the fibre length. Moreover, one could also exploit the intrinsic random nature of the multimode coupling and output speckle patterns to securely encode and authenticate data as proposed recently \cite{Goorden:14,random} where the transmission fibre itself would play the role of the encoding medium, with possible extensions also to the control of quantum states for quantum sensing and simulation \cite{quantum}. \\

\noindent {\bf{Data availability}}\\
An example of the code is provided along with data for 1m at http://dx.doi.org/10.5525/gla.researchdata.751.\\

\noindent {\bf{Code availability}}\\
The code has been presented and explained in the Supplementary Note 4: Software. Furthermore, a code example is available at the DOI link. \\

\noindent {\bf{Acknowledgements}}\\
DF and RMS acknowledge financial support from EPSRC (UK, grant no. EP/M01326X/1).\\




\begin{appendix} 
\onecolumngrid

\newpage
\lstset{language=Python}
\lstset{style=python-idle-code}

\begin{center}{
	\noindent {\large{\bf{Transmission of natural scene images through a multimode fibre: Supplementary Material}}}\\
	\vskip 4mm

	\noindent{Piergiorgio Caramazza\textsuperscript{1}, Ois{\'i}n Moran\textsuperscript{2},  Roderick Murray-Smith\textsuperscript{2,*}, Daniele Faccio\textsuperscript{1,*}}\\
	\vskip 0.5mm
	\noindent{\it{\small{$^1$School of Physics and Astronomy, University of Glasgow, Glasgow, G12 8QQ, UK\\
$^2$School of Computing Science, University of Glasgow, Glasgow, G12 8QQ, UK 
      }}}\\
	\noindent{\footnotesize{\textsuperscript{*}Email: Daniele.Faccio@glasgow.ac.uk, Roderick.Murray-Smith@glasgow.ac.uk.}}\\
	\vskip 4mm
	\noindent \small{Supplementary information to ``Transmission of natural scene images through a multimode fibre'' \\providing additional details regarding the code and measurements.}\\
	\vskip 12mm
     } 
\end{center}

\renewcommand{\figurename}{\bf{Supplementary Figure}}
\setcounter{figure}{0}
%
%
%
%
%
%
%
%
%
%
%
\begin{center}{
	\noindent {\uppercase{\bf{\small{Supplementary Note 1: Effect of image size at fibre input}}}}}\\ 
\end{center}
 \begin{figure*}[b!]
	{\includegraphics[width=16cm]{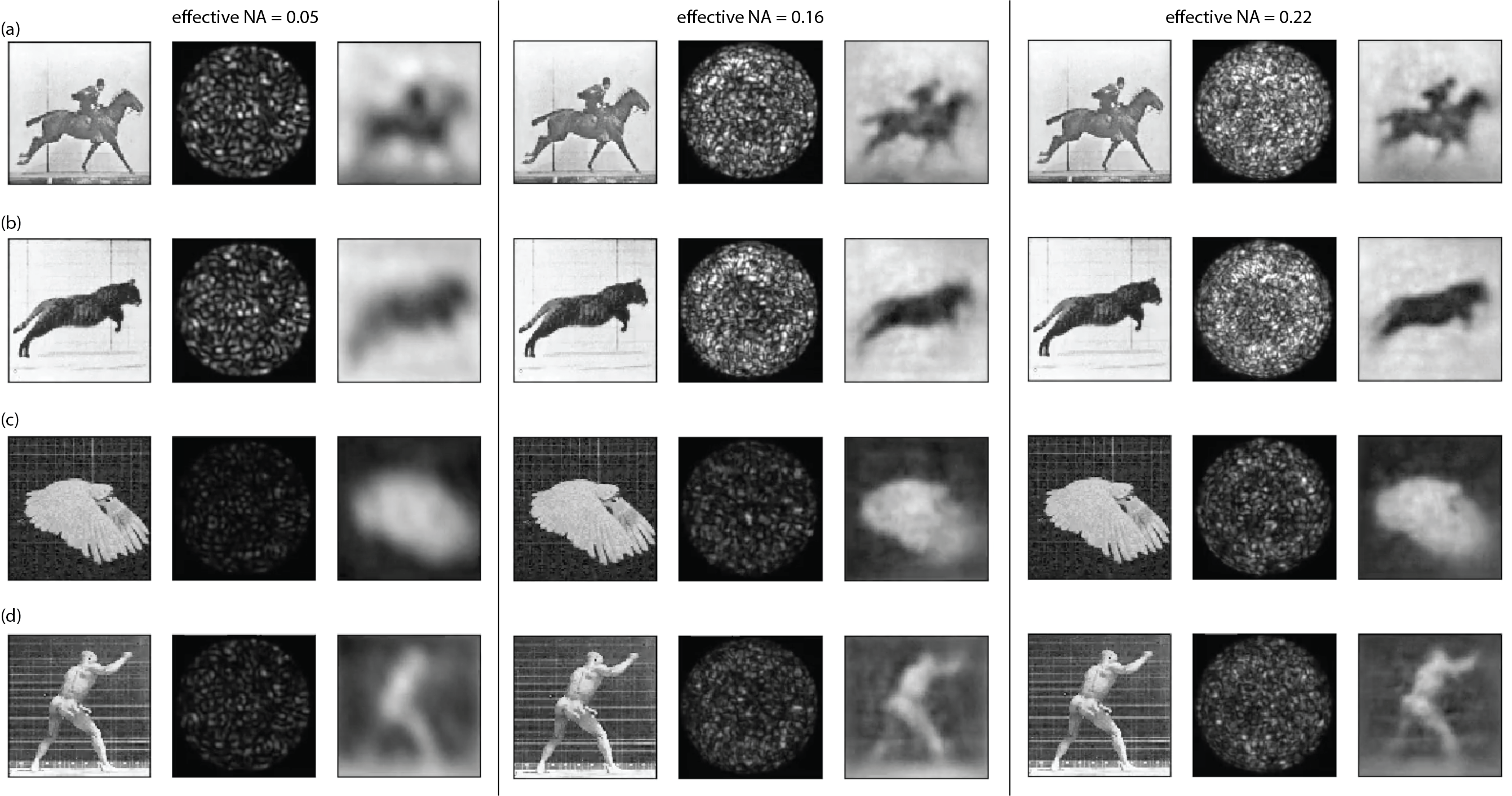}}
	\caption{ \normalfont{Comparison between same images focused with different angular spread reported by an estimation of the effective NA (this was realized by putting a telescope before the objective with different magnification). Each image (a)-(d) is reported along with its relative speckle pattern at the fiber output and ANN reconstruction. An increase in the effective NA leads to averaged smaller speckles (indicating that higher spatial frequency modes have been excited). } }
	\label{compare}
\end{figure*}
Whilst performing experiments we investigated the effect of different focusing (i.e. image size) configurations at the fibre input. We performed three different experiments in which we used exactly the same data/images for the retrieval of the $W$ matrix and also the same test images. However, by inserting an additional telescope straight after the SLM, we could control the images sizes before the objective coupling into the fibre.  {By demagnifying the input images by factors 1x, 1.4x and 4x, we expect the angular spread with which the images are being focused to decrease. In these three cases, an effective numerical aperture can be estimated by considering the objective focal length ($f = 34$ mm) and the image size at the input of the fiber. Respectively, the sides of the images are: $10.5$ mm, $7.5$ mm and  $2.6$ mm, corresponding to NA$_{eff}=0.22 $, NA$_{eff} = 0.16$ and NA$_{eff} = 0.05$}. In all three case, the images were placed in a similar position on the fibre, i.e. slightly off to the centre. Indeed, we noticed that placing the image directly in the centre of the fibre led to very clear accumulation of the speckles patterns towards the centre of the fibre, surrounded by broken ring-like structures in the outer region. These ring-like structures are a clear indication of higher modes, in keeping with previous reports also from other groups [1]. However, the fact that these modes maintain their ring-like structure as opposed to a more random speckle structure indicates that only a few of the higher order modes have been excited. Conversely, the full speckle pattern, distributed across the full fibre output, is obtained only by exciting many modes. This condition was observed by displacing the beam slightly to either side with respect to the central position. This is the desired configuration as the objective here is to image with the highest resolution possible: the fine features of any image are carried by the higher spatial frequencies, which in turn correspond to the higher order modes in the fibre.\\
The effect of changing the inout  focusing condition can be clearly seen in Fig.~\ref{compare}. When focusing with the {smallest NA$_{eff}$} at the fibre input facet, the speckles at the output are largest, corresponding to fewer modes and the final retrieval is significantly worse when compared to the {largest NA$_{eff}$}. The intermediate { NA$_{eff}$} shows slightly worse results with respect to the {largest NA$_{eff}$}. A consistent trend is therefore found between the  {effective NA}, the size and number of speckles at the fibre output and final image quality.\\

\vskip 7 mm
\begin{center}{
	\noindent {\uppercase{\bf{\small{Supplementary Note 2: Methods for image comparison}}}}}\\ 
\end{center}
\vskip 1.7 mm
Two methods have been considered in order to quantify the quality of our predicted images: the structural similarity index (SSIM) [2] and the Parson correlation coefficient (PCC). In both cases, a perfect match would correspond to the maximum value 1. Considering two images $X$ and $Y$, we use the definitions:

\begin{equation}
SSIM(X,Y) = \frac{(2\mu_X \mu_Y + C_1)(2\sigma_{XY}+C_2)}{(\mu_X^2 \mu_Y^2 + C_1)(\sigma_X^2\sigma_Y^2+C_2)}
\end{equation}

\noindent where $\mu_{X}$ represent the average of $X$, $\mu_{Y}$ the average of $Y$, $\sigma_{XY}$ the covariance of $X$ and $Y$, $\sigma_{X}^2$  the variance of $X$ and $\sigma_Y^2$  the variance of $Y$. Whereas, $C_{1}$ and $C_2$ are two parament defined as $C_1 = (K_1 L)**2$ and $C_1 = (K_2 L)**2$  where $K_1$ was set to $0.01$ and $K_2$ to $0.03$ and $L$ is the dynamic range of the image pixels. Instead, the Parson correlation coefficent is defined as:

\begin{equation}
PCC(X,Y) = \frac{\sum_i (x_i -\bar X)(y_i -\bar Y)}{\sqrt{\sum_i (x_i -\bar X)^2\sum_i (y_i -\bar Y)^2} }
\end{equation}

\noindent where $x_i$ and $y_i$ indicate the pixels with index $i$  respectively of the images $X$ and $Y$, and $\bar X$ (similarly with $Y$) the average of $X$.

\vskip 7 mm
\begin{center}{
	\noindent {\uppercase{\bf{\small{Supplementary Note 3: Data}}}}}\\ 
\end{center}
\vskip 1.7 mm

For optimisation and testing of the model parameters, we use 50,000 images from the ImageNet collection [3]. {The training and experimental datasets are supplied as additional material [4]}.
 \begin{figure*}[t!]
	{\includegraphics[width=16cm]{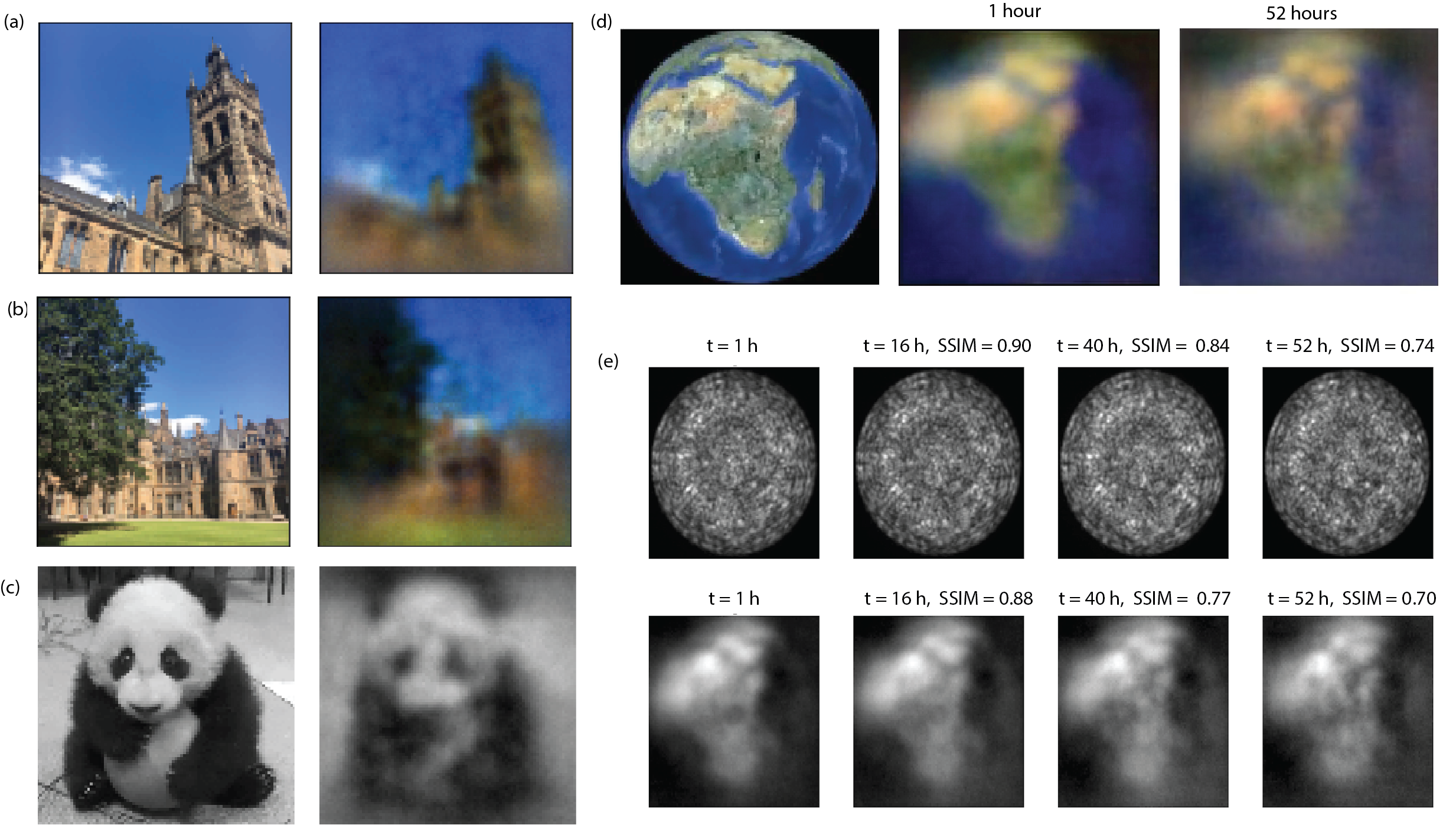}}
	\caption{\normalfont{ Additional examples of natural scenes imaged through a mulitmode fiber. (a)-(b) Colour photographs of Glasgow University (imaged through a 1 meter fibre). (c) Black and white photograph of a panda (imaged through 10 m fibre). (d) Satellite photograph of Earth imaged at two different times after the transmission of the initial training dataset (1 hour and 52 hours, imaged through 10 m fibre). {(e) Comparison between green-colour-channel speckles of the images in (d), at different times. The speckle patterns at time 1 hour, 16 hours, 40 hours and 52 hours are shown on the first column together with the SSIM parameters calculated respect to the time ``1 hour'' speckle pattern. As can be seen, the 10 m length of the fibre leads to a slow decorrelation of the speckle pattern, quantified here by the decreasing SSIM with time. Similarly, the reconstructed green-channel-images show a slow degradation of the SSIM with a linear dependence with respect to the speckle SSIM degradation.}}  }
	\label{other_examples}
\end{figure*}
Validation image examples are images and videos from the Muybridge collection such as a running horse, a jumping cat and a flying parrot. 
As explained in the main text, the behaviour of the image retrieval seems to be largely independent of the actual test images that was chosen. We focused mainly on the Muybridge images in the main text but here show other examples in Fig.~\ref{other_examples}.  Figures~\ref{other_examples}(a) and (b) show colour photographs of the University of Glasgow, imaged through a 1 m fibre. (c) is a grayscale image of a panda. (d) is a satellite image of the Earth imaged through a 10 m fibre at a two different times after the initial training an inversion process is completed showing that the retrieved inversion matrix and setup can still be used at a later time, as would be expected (yet still need to be verified) for a robust inversion system. {In (e) we report the speckle patterns relative to (d) at different times (1 hour, 16 hours, 40 hours and 52 hours) of a single channel of the RGB image. In order to allow a quantitative comparison, we used the SSIM defined in the previous section.}  We note that the setup was not placed in a specifically engineered or stabilised environment and was thus subject to standard night-day temperature fluctuations (2-3 degrees) and environmental vibrations. {On the other hand, as we can see in Fig.~\ref{other_examples}-(e), the correlation between the speckle pattern in time is still high, allowing a good reconstruction. We judged out the scope of the present work to introduce models able to deal with deep changes in the system transmission matrix, such as in presence of relevant bending or temperature variations. Indeed, in future work it would be interesting to explore the possibilities given by a physics inspired artificial neural network also respect to these challenges.}\\
In Fig.~\ref{image_collection} we show a collection of images taken from the ImageNet database together with their respective output speckle patterns (at the output of 1 m long fibre) and final reconstructions. These images provide further evidence for the image variability and robustness of the imaging reconstruction.

\vskip 12 mm
\begin{center}{
	\noindent  {\uppercase{\bf{\small{Supplementary Note  4: Software}}}}}\\ 
\end{center}
\vskip 1.7 mm
The code was developed in Python 3.6.5 with a standard Anaconda \url{http://www.anaconda.com/download} configuration, including  Keras [5] and TensorFlow [6]. The code is supplied as additional supplementary material that can be downloaded together with all of the training and experimental data/images [4].



%

\vskip 5 mm
\begin{center}{
	\noindent {\bf{A. Model specification}}}\\ 
\end{center}

\vskip 1 mm
The model is implemented as a simple complex, densely connected layer. The individual weights are regularised with an $L_2$ minimising term, weighted by $\lambda=0.03$. weights are initialised randomly, uniformly between $\pm 0.002$. Recorded images are collected into a training set of $N=45,000$ and validation set of $5,000$.  

\begin{lstlisting}[caption=Model specification]
speckle_dim = 120
out_dim = 92
lamb = 0.03

inp = Input(shape=(speckle_dim**2,2))
comp = ComplexDense(out_dim**2, use_bias=False, 
		    kernel_initializer=RandomUniform(-.002, .002),
		    kernel_regularizer=regularizers.l2(lamb))(inp)
amp = Amplitude()(comp)

model = Model(inputs=inp, outputs=amp)
\end{lstlisting}
The ComplexDense Layer is a custom layer we developed for Keras. It is a straightforward Dense layer, but with complex-valued weights. The complex weights are represented with Complex64 64-bit data types. Its only task is to implement the complex-valued multiplication.
\begin{lstlisting}[caption=Custom ComplexDense Layer]
class ComplexDense(Layer):

def __init__(self, output_dim,
             activation=None,
             use_bias=True,
           	 kernel_initializer='glorot_uniform',
             bias_initializer='zeros',
             kernel_regularizer=None,
             **kwargs):
	super(ComplexDense, self).__init__(**kwargs)
    self.output_dim = output_dim
    self.activation = activations.get(activation)
    self.use_bias = use_bias
    self.kernel_initializer = initializers.get(kernel_initializer)
    self.bias_initializer = initializers.get(bias_initializer)
    self.kernel_regularizer = regularizers.get(kernel_regularizer)
        

def build(self, input_shape):
    self.kernel = self.add_weight(name='kernel',
                                  shape=(input_shape[1], self.output_dim, 2),
                                  initializer=self.kernel_initializer,
                                  regularizer=self.kernel_regularizer,
                                  trainable=True)

    if self.use_bias:
        self.bias = self.add_weight(name='bias',
                                   shape=(self.output_dim, 2),
                                   initializer=self.bias_initializer,
                                   trainable=True)
    else:
        self.bias = None
    super(ComplexDense, self).build(input_shape)

def call(self, X):
    # True Complex Multiplication (by channel combination)
    complex_X = channels_to_complex(X)
    complex_W = channels_to_complex(self.kernel)

    complex_res = complex_X @ complex_W
        
    if self.use_bias:
        complex_b = channels_to_complex(self.bias)
             = K.bias_add(complex_res, complex_b)
        
    output = complex_to_channels(complex_res)

    if self.activation is not None:
        output = self.activation(output)

    return output

def compute_output_shape(self, input_shape):
    return (input_shape[0], self.output_dim, 2)

def get_config(self):
    config = {'output_dim': self.output_dim,
	     'use_bias': self.use_bias,
	     'kernel_initializer': initializers.serialize(self.kernel_initializer),
	     'bias_initializer': initializers.serialize(self.bias_initializer),
	     'kernel_regularizer': regularizers.serialize(self.kernel_regularizer)
	      }
    base_config = super(ComplexDense, self).get_config()
    return dict(list(base_config.items()) + list(config.items()))

\end{lstlisting}

\vskip 5 mm
\begin{center}{
	\noindent {\bf{B. Parameter optimisation}}}\\ 
\end{center}

\vskip 1 mm
The model fitting uses the standard Keras routines, involving a choice of stochastic gradient descent and mean square error for the cost function. Here the data variable \textit{x\_train, y\_train, x\_validation, y\_validation, x\_text and y\_test} refer to the amplitudes of, respectively, speckle patterns (\textit{x}) and original images (\textit{y}). 
\begin{lstlisting}[caption=Compile and optimise parameters]
model.compile(optimizer=SGD(lr=1e-5), loss='mse', metrics=['mse'])
model_chk = ModelCheckpoint(weights_filepath, monitor='mse', verbose=0, 
			    save_best_only=False, 
	 		    save_weights_only=False, mode='auto', period=1)
reduce_lr = ReduceLROnPlateau(monitor='loss', factor=0.1, patience=2,
	    min_lr=lr/1e3, verbose=1,)
early_stop = EarlyStopping(monitor='loss', min_delta=0.0001, patience=8)

model.fit(x_train, y_train, validation_data = (x_validation, y_validation),
	  epochs = 850, batch_size = 32, 
	  callbacks = [model_chk, reduce_lr, early_stop], shuffle = True)
\end{lstlisting}

Once the network parameters have converged (we ran the network for 850 iterations which takes ca 2 days on a PC with Nvidia TitanXp GPU card, you can generate predictions of outputs using 
\begin{lstlisting}
pred_test = model.predict(x_test)**2
\end{lstlisting}

\vskip 11 mm
\begin{center}{
	\noindent {\uppercase{\bf{\small{Supplementary References}}}}}
\end{center}
\vskip -1pt

\expandafter\ifx\csname url\endcsname\relax
  \def\url#1{\texttt{#1}}\fi
\expandafter\ifx\csname urlprefix\endcsname\relax\def\urlprefix{URL }\fi
\providecommand{\bibinfo}[2]{#2}
\providecommand{\eprint}[2][]{\url{#2}}

\begin{enumerate}[{label=[\arabic{*}]}]
\itemsep-0.4em
\begin{small}
\item Papadopoulos, I.~N., Farahi, S., Moser, C. \& Psaltis, D Focusing and scanning light through a multimode
  optical fiber using digital phase conjugation. \textit{Optics Express} \textbf{20}, 10583--10590 (2012).
\item Wang, Z., et~al. {Image quality assessment: from error visibility to structural similarity.} \textit{IEEE transactions on image processing} \textbf{13}, 600--612 (2004).

\item Deng, J., et~al. {ImageNet: A Large-Scale Hierarchical Image Database.} In \textit{CVPR09} (2009).

\item {Caramazza, P.}, {Moran, O.}, Murray-Smith, R.  \& {Faccio, D.} \textit{Data can be downloaded from the University of
   Glasgow repository:}  \urlprefix\url{DOI: http://dx.doi.org/10.5525/gla.researchdata.751}

\item {Chollet, F.}, {et~al.} Keras.  \urlprefix\url{https://keras.io} (2015).

\item {Abadi, M.} {et~al.} {TensorFlow}: Large-scale machine learning on
 heterogeneous systems ({2015}).
 \urlprefix\url{https://www.tensorflow.org/}. {Software available from tensorflow.org}.
\end{small}
\end{enumerate}

\newpage

\begin{figure*}[t!]
	\includegraphics[width=16cm]{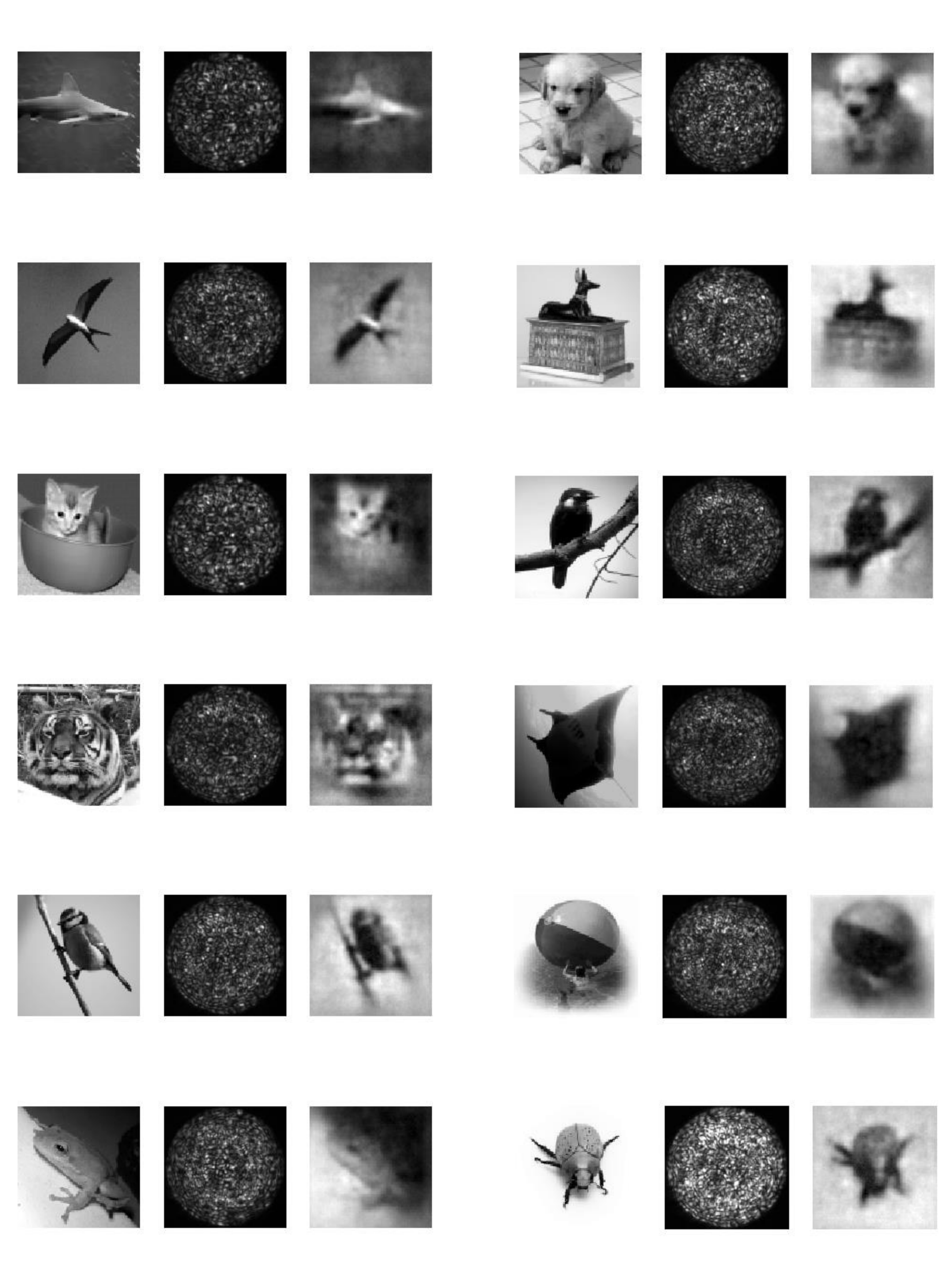}
	\caption{\normalfont{A sample collection of images from the ImageNet database, as projected into the fibre together with their relative output speckle patterns and retrieved images. {These testing images were not present in the training dataset.}}}
	\label{image_collection}
\end{figure*}

\newpage


%
%
%

\end{appendix}


\end{document}